%% file: root.tex
\documentclass[letterpaper, 10 pt, conference]{ieeeconf}  

\IEEEoverridecommandlockouts                              

\usepackage{amsmath}
\usepackage{amsfonts}
\usepackage{amssymb}
\usepackage{soul,xcolor}
\usepackage{graphicx}
\usepackage{subfigure}
\usepackage{epsfig} 
\usepackage{epstopdf}
\usepackage{multicol}
\usepackage{mathtools}
\usepackage{mathrsfs}
\usepackage{multirow}
\usepackage{comment}
\usepackage{array}
\usepackage{empheq}
\usepackage{balance}
\usepackage{threeparttable}
\usepackage{cite}
\newtheorem{theorem}{Theorem}   
\newtheorem{lemma}{Lemma}

\newtheorem{remark}{Remark}

\newtheorem{definition}{Definition}
\newcommand{\tr}{\mathsf{ T}}

\title{\LARGE \bf
	Dimension Reduction for Efficient Data-Enabled Predictive Control
}

\author{Kaixiang~Zhang, Yang~Zheng, Chao~Shang, and Zhaojian~Li$^{*}$ 
	\thanks{\quad $*$Zhaojian Li is the corresponding author.}
	\thanks{Kaixiang Zhang, Zhaojian Li are with the Department of Mechanical Engineering, Michigan State University, East Lansing, MI 48824, USA (e-mail: \{zhangk64, lizhaoj1\}@msu.edu).}
	\thanks{Yang Zheng is with the Department of Electrical and Computer Engineering, University of California San Diego, CA 92093, USA, (e-mail: zhengy@eng.ucsd.edu).}
	\thanks{Chao Shang is with the Department of Automation, Tsinghua University, Beijing 100084, China, (e-mail: c-shang@tsinghua.edu.cn).}
}

\begin{document}

\maketitle
\thispagestyle{empty}
\pagestyle{empty}

\begin{abstract}	
	In this letter, we propose a simple yet effective singular value decomposition (SVD) based strategy to reduce the optimization problem dimension in data-enabled predictive control (DeePC). Specifically, in the case of linear time-invariant systems, the excessive input/output measurements can be rearranged into a smaller data library for the non-parametric representation of system behavior. Based on this observation, we develop an SVD-based strategy to pre-process the offline data that achieves dimension reduction in DeePC. Numerical experiments confirm that the proposed method significantly enhances the computation efficiency without sacrificing the control performance.
\end{abstract}
	
\section{Introduction}
With advancement in computing and sensing technologies, data-driven control approaches have become increasingly prevalent in the context of complex dynamic systems \cite{Markovsky2021ARC,Persis2020TAC}. When modeling based on first-principles is infeasible or too costly, data-driven control approaches can circumvent explicit system models and directly incorporate collected data for control development, offering an appealing alternative to classical model-based methods \cite{Markovsky2021ARC}.

Recent developments in data-driven model predictive control (MPC) have shown great promise to achieve optimal control with simultaneous constraint satisfaction and stability guarantees~\cite{Rosolia2018TAC,Hewing2020AR}. Particularly, Data-EnablEd Predictive Control (DeePC)~\cite{Coulson2019ECC} receives increasing attention as it can directly exploit input/output data to achieve safe and optimal control of unknown systems. In contrast to conventional MPC schemes that depend on explicit parametric models, DeePC leverages Willems' fundamental lemma~\cite{Willems2005SCL} to represent system trajectories in a non-parametric manner. Specifically, the fundamental lemma shows that a data Hankel matrix consisting of a pre-collected input/output trajectory of a linear system spans the vector space of all trajectories that the system can produce, given that the input sequence is persistently exciting~\cite{Willems2005SCL}. This lemma plays an essential role in DeePC and has been applied to tackle various control problems \cite{Persis2020TAC,Berberich2020TAC,Coulson2021TAC,Baros2022AUTO}; see~\cite{Markovsky2021ARC} for an extensive review. DeePC has found success across diverse applications, including quadcopters~\cite{Elokda2021IJRNC}, power systems~\cite{Huang2021TCST,Mahdavipour2022IFAC}, and connected and autonomous vehicles~\cite{Wang2022arXiv,wang2022experimental}.

All the aforementioned applications show the value of the fundamental lemma and the potential of the DeePC method. Yet, a major challenge in DeePC is the high computational complexity; at each iteration, it needs to solve an optimization problem with high-dimensional variables, which might be computationally intractable in resource-limited situations. In particular, to implement DeePC, sufficiently rich input/output data should be recorded offline, and then the Hankel matrix is used to store the data for the non-parametric representation of unknown systems. Since the dimension of optimization variables in DeePC is determined by the length of pre-collected data, excessive up-front data collection will lead~to overly high optimization dimensions in the subsequent online calculation. For real-time implementation in resource-limited situations, it is necessary to reformulate (and possibly approximate) the original optimization problem in DeePC to reduce the dimension of optimization variables for efficient computation.

In this letter, we introduce a \textit{minimum-dimension} version of the DeePC method, which admits significantly better computational efficiency with no/little degradation in control performance. The idea is based on a well-known observation that the Hankel matrix in the fundamental lemma is always \textit{low-rank}. Thus, a new data matrix with a smaller column dimension could be constructed to represent the system input/output behavior, which balances the tradeoff between control performance and computational complexity. Based on this observation, we develop a singular value decomposition (SVD) based strategy to extract the principal components from a large data library (i.e., the Hankel matrix). The resulting smaller data library is then incorporated into DeePC, which effectively reduces the dimension of optimization variables. We denote this new formulation as minimum-dimension DeePC, since the dimension of optimization variables becomes minimal to represent the system behavior.

We note that the SVD-based technique was utilized in~\cite{YangASCC2015} as a heuristic to reduce the dimension of the Hankel matrix, but there was no analysis to justify the design. In this letter, the rationale behind using SVD for dimension reduction is formalized, and the conclusions are generalizable to other forms of data matrices (e.g., Page matrix and mosaic-Hankel matrix). In addition, similar SVD-based strategies have been used in data-driven control~\cite{Markovsky2021ARC,Coulson2019ECC,Huang2021TCST} and system identification~\cite{Zeiger1974TAC,Damen1982SCL} but with different purposes.
Specifically, the studies~\cite{Markovsky2021ARC,Coulson2019ECC,Huang2021TCST,Zeiger1974TAC,Damen1982SCL} use SVD to get a low-rank matrix approximation for the sake of denoising, while we aim to attain reduced-dimensional problems for the purpose of efficient computation. Thus, the Hankel matrix after the SVD pre-processing in~\cite{Markovsky2021ARC,Coulson2019ECC,Huang2021TCST,Zeiger1974TAC,Damen1982SCL} still remains the same size, while our SVD-based strategy reduces the dimension of the Hankel matrix significantly. Extensive simulations validate the performance of our SVD-based strategy for DeePC and confirm the benefits of improving numerical efficiency without compromising the control performance.
	
The remainder of this letter is organized as follows. Section~\ref{sec_preliminaries} introduces the non-parametric representation of linear systems and the DeePC paradigm. Section~\ref{sec_svd} presents our observation on the Hankel data rearrangement and the SVD-based strategy for dimension reduction. Simulations are shown in Section~\ref{sec_simulation}. Finally, we conclude the letter in Section~\ref{sec_conclusion}.
	
\textit{Notation:} 
We use $0_{n\times m}$ and $I_{n}$ to denote a zero matrix of size $n\times m$ and an $n\times n$ identity matrix, respectively. Given a signal $\omega(t) \in \mathbb{R}^{n}$ and two integers $i, j\in \mathbb{Z}$ with $i\le j$, we denote by $\omega_{\left[i, j\right]}$ the restriction of $\omega(t) \in \mathbb{R}^n$ to the interval $\left[i, j\right]$, namely, $\omega_{\left[i, j\right]} := \begin{bmatrix}
	\omega^{\tr}(i), \omega^{\tr}(i+1), \cdots, \omega^{\tr}(j)
\end{bmatrix}^{\tr}$. To simplify notation, we also use $\omega_{\left[i, j\right]}$ to denote the sequence $\left\lbrace \omega(i), \cdots, \omega(j)\right\rbrace$. The Hankel matrix of depth $k \in \mathbb{Z}$ ($k\le j-i+1$) associated with $\omega_{\left[i, j\right]}$ is defined as
\abovedisplayskip= 4pt 
\belowdisplayskip= 4pt
\[
{\small
	\mathcal{H}_{k}(\omega_{\left[i, j\right]}):= \begingroup 
	\setlength\arraycolsep{2pt}
	\begin{bmatrix}
		\omega(i) & \omega(i+1) & \cdots & \omega(j-k+1) \\
		\omega(i+1) & \omega(i+2) & \cdots & \omega(j-k+2)\\
		\vdots & \vdots & \ddots & \vdots \\
		\omega(i+k-1) & \omega(i+k) & \cdots & \omega(j)
	\end{bmatrix}.
	\endgroup
}
\]
\begin{definition}
	The sequence $\omega_{[i,j]}$ is said to be \textit{persistently exciting of order $k$} 
	if $\mathcal{H}_{k}(\omega_{[i,j]})$ has full row rank of $nk$. 
\end{definition}
	
\section{Preliminaries}\label{sec_preliminaries}
In this section, we overview a non-parametric representation of linear systems~\cite{Willems2005SCL} and the DeePC formulation~\cite{Coulson2019ECC}.

\subsection{Non-Parametric Representation of Linear Systems} 
Consider a discrete-time linear time-invariant (LTI) system 
\begin{equation} \label{equ_LTI}
	{
		\begin{aligned}
			x(t+1) &= Ax(t) + Bu(t), \\
			y(t) &= Cx(t) + D u(t),
		\end{aligned}
	}
\end{equation}
where $A\in \mathbb{R}^{n\times n}$, $B\in \mathbb{R}^{n\times m}$, $C\in \mathbb{R}^{p\times n}$, $D\in \mathbb{R}^{p\times m}$ are system matrices, and $x(t) \in \mathbb{R}^n$, $u(t) \in \mathbb{R}^m$, and $y(t) \in \mathbb{R}^p$ denote the state, control input, and output, respectively. 

Model~\eqref{equ_LTI} is a parametric description of the system, defined by $(A, B, C, D)$. Willems' fundamental lemma allows us to represent~\eqref{equ_LTI} via a finite collection of its input/output data. Let $(u^{\mathrm{d}}_{[0,T-1]}$, $y^{\mathrm{d}}_{[0,T-1]})$ be a length-$T$ 
input/output trajectory of system \eqref{equ_LTI}. The Hankel matrices $\mathcal{H}_{L}(u^{\mathrm{d}}_{[0,T-1]})$ and $\mathcal{H}_{L}(y^{\mathrm{d}}_{[0,T-1]})$ 
are given by
\begin{equation} \label{equ_inputoutputHankel}
	{\small
		\begin{bmatrix}
			\mathcal{H}_{L}(u^{\mathrm{d}}_{[0,T-1]}) \\ \hline
			\mathcal{H}_{L}(y^{\mathrm{d}}_{[0,T-1]})
		\end{bmatrix} \!:= \! \begingroup 
		\setlength\arraycolsep{2pt}
		\begin{bmatrix} u^{\mathrm{d}}(0) & u^{\mathrm{d}}(1) & \cdots & u^{\mathrm{d}}(T\!-\!L) \\
			\vdots & \vdots & & \vdots \\
			u^{\mathrm{d}}(L\!-\!1) & u^{\mathrm{d}}(L) & \cdots & u^{\mathrm{d}}(T\!-\!1) \\ \hline
			y^{\mathrm{d}}(0) & y^{\mathrm{d}}(1) & \cdots & y^{\mathrm{d}}(T\!-\!L) \\
			\vdots & \vdots & & \vdots \\
			y^{\mathrm{d}}(L\!-\!1) & y^{\mathrm{d}}(L) & \cdots & y^{\mathrm{d}}(T\!-\!1)
		\end{bmatrix}\!,
		\endgroup
	}
\end{equation}
where each column is a length-$L$ input/output trajectory~of~\eqref{equ_LTI}. The column space gives a non-parametric representation of \eqref{equ_LTI} when the input sequence is persistently exciting as shown in the following lemma.

\begin{lemma}[Fundamental Lemma \cite{Persis2020TAC,Willems2005SCL}]\label{lemma_Willems}
	Consider a controllable LTI system \eqref{equ_LTI} and assume that the input sequence $u^{\mathrm{d}}_{[0,T-1]}$ is persistently exciting of order $n+L$. Then, any length-$L$ sequence  $(u_{[0, L-1]}, y_{[0,L-1]})$ is an input/output trajectory of \eqref{equ_LTI} if and only if we have
	\begin{equation} \label{eq:fundamental-lemma}
		{
			\begin{bmatrix}
				u_{[0,L-1]} \\ y_{[0,L-1]}
			\end{bmatrix} = \begin{bmatrix}
				\mathcal{H}_{L}(u^{\mathrm{d}}_{[0,T-1]}) \\
				\mathcal{H}_{L}(y^{\mathrm{d}}_{[0,T-1]})
			\end{bmatrix} g
		}
	\end{equation}
	for some real vector $g \in \mathbb{R}^{T-L+1}$.
\end{lemma}

\subsection{Data-EnablEd Predictive Control}
Conventional control strategies rely on the explicit system model $(A, B, C, D)$ in~\eqref{equ_LTI} to facilitate the controller design, while DeePC~\cite{Coulson2019ECC}, \cite[Section 5.2]{Markovsky2021ARC} is a non-parametric approach which bypasses system identification and directly utilizes pre-collected input/output data to design a safe control policy. In particular, DeePC employs pre-collected data to predict the system behavior based on the fundamental lemma.

Let $T_{\mathrm{ini}}$, $N\in \mathbb{Z}$, and $L = T_{\mathrm{ini}} + N$. We choose a sufficiently long input sequence $u^{\mathrm{d}}_{[0,T-1]}$ of length $T$, which is persistently exciting of order $n+L$. Let $y^{\mathrm{d}}_{[0,T-1]}$ be the corresponding output sequence. We divide the Hankel~matrices $\mathcal{H}_{L}(u^{\mathrm{d}}_{[0,T-1]})$ and $\mathcal{H}_{L}(y^{\mathrm{d}}_{[0,T-1]})$ into the two parts (i.e., ``past data'' of length $T_{\mathrm{ini}}$ and ``future data'' of length $N$):
\begin{equation}\label{eq:Hankel}
	\begin{aligned}
		\begin{bmatrix}U_{\mathrm{p}}\\{U}_{\mathrm{f}}
		\end{bmatrix} = \mathcal{H}_{L}({u^{\mathrm{d}}_{[0,T-1]}}), \quad
		\begin{bmatrix}{Y}_{\mathrm{p}}\\{Y}_{\mathrm{f}}
		\end{bmatrix} = \mathcal{H}_{L}({y^{\mathrm{d}}_{[0,T-1]}}),
	\end{aligned}
\end{equation}
where $U_{\mathrm{p}}$ and $U_{\mathrm{f}}$ denote the first $T_{\mathrm{ini}}$ block rows and the last $N$ block rows of $\mathcal{H}_{L}({u^{\mathrm{d}}_{[0,T-1]}})$, respectively (similarly for $Y_{\mathrm{p}}$ and $Y_{\mathrm{f}}$). Let $u_{\mathrm{ini}}=u_{\left[t-T_{\mathrm{ini}}, t-1\right]}$ be the control input sequence within a past time horizon of length $T_{\mathrm{ini}}$, and $u= u_{\left[t, t+N-1\right]}$ be the control input sequence within a prediction horizon of length $N$ (similarly for $y_\mathrm{ini}$ and $y$). The DeePC solves the following constrained optimization problem at time step $t$:
\begin{align}  \label{equ_DeePC}
	\min_{g,u,y,\sigma_{u},\sigma_{y}} & \left\|y - y_r\right\|_{Q}^{2}+\left\|u\right\|_{R}^{2}
	+ \lambda_{u}\left\|\sigma_{u} \right\|_{2}^{2} \nonumber \\
	&\qquad \qquad \qquad \qquad  + \lambda_{y}\left\|\sigma_{y} \right\|_{2}^{2} + \lambda_{g}\left\|g \right\|_{2}^{2} \nonumber 
	\\
	\mathrm{subject~to} &\,\,  \begin{bmatrix}U_{\mathrm{p}}\\{U}_{\mathrm{f}}\\{Y}_{\mathrm{p}}\\{Y}_{\mathrm{f}}
	\end{bmatrix}g = \begin{bmatrix}
		u_{\mathrm{ini}} \\ u \\ y_{\mathrm{ini}} \\ y
	\end{bmatrix} + \begin{bmatrix}
		\sigma_{u} \\ 0 \\ \sigma_{y} \\ 0
	\end{bmatrix},
	u \in\mathcal{U}, y \in\mathcal{Y},
\end{align}
where $y_{r} = \begin{bmatrix}
	y_{r}^{\tr}(t), y_{r}^{\tr}(t+1), \cdots, y_{r}^{\tr}(t+N-1)
\end{bmatrix}^{\tr}$ is a reference trajectory, $Q\in \mathbb{S}^{pN}_+$, $R\in \mathbb{S}^{mN}_+$ are weighting matrices, $\mathcal{U}$, $\mathcal{Y}$ represent the input and output constraints, respectively, $\sigma_{u}\in \mathbb{R}^{mT_{\mathrm{ini}}}$, $\sigma_{y} \in \mathbb{R}^{pT_{\mathrm{ini}}}$ are auxiliary variables, and $\lambda_{u} \ge 0$, $\lambda_{y} \ge 0$, $\lambda_{g} \ge 0$ are regularization parameters.

DeePC solves \eqref{equ_DeePC} in a receding horizon fashion. After computing the optimal sequence $u^{*}=\begin{bmatrix}
	u_{0}^{*\tr} & \cdots & u_{N-1}^{*\tr}
\end{bmatrix}^{\tr}$, we apply $(u(t), \cdots, u(t+l-1)) = (u_{0}^{*}, \cdots, u_{l-1}^{*})$ to the system for some $l< N$ steps. Then, when time is shifted to $t+l$, $(u_{\mathrm{ini}}, y_{\mathrm{ini}})$ is updated to the most recent input/output data; see \cite{Coulson2019ECC,Coulson2021TAC} for more details. 

\begin{remark}[Dimension of $g$~in~DeePC]  \label{remark-1}
	To guarantee the persistent excitation of the sequence $u^{\mathrm{d}}_{[0,T-1]}$, the column number of the Hankel matrix $\mathcal{H}_{n+L}(u^{\mathrm{d}}_{[0,T-1]})$ must be at least no less than its row number. This implies that the number of data points $T$ must at least satisfy $T - (n + L) + 1 \geq m(n+L)$, i.e., $T \geq (m+1)(n+L) -1$.
	Therefore, the dimension of $g$ in \eqref{equ_DeePC} is lower bounded as
	\begin{equation} \label{eq:g-dimension}
		{\small
			T-L+1 \geq mL + (m+1)n,
		}
	\end{equation}
	where $L = T_{\mathrm{ini}} + N$. Thus, a large $T$ leads to a high dimension of the optimization variable $g$ in \eqref{equ_DeePC}, which increases the computation burden and thus hinders the deployment of DeePC in resource-limited situations.
\end{remark}

\begin{remark}[Practical choice of $T$] \label{remark-2}
	In~\eqref{equ_DeePC}, the value of $T_{\mathrm{ini}}$ needs to be larger than the observability index\footnote{The smallest integer $l$ such that
		$\begin{bmatrix}C^\tr, (CA)^\tr, \ldots, (CA^{l-1})^\tr\end{bmatrix}^\tr$ has full column rank $n$.} in order to estimate the system initial state $x_{\mathrm{ini}}$ at time step $t$~\cite[Lemma 4.1]{Coulson2019ECC}, \cite[Lemma 1]{Markovsky2021ARC}. This observability index (upper bounded by $n$) and the system internal state dimension $n$ may be unknown when the system model~\eqref{equ_LTI} is unknown. In practical applications, one would need to collect a sufficiently large amount of data points to satisfy $T \geq (m+1)(n+L) -1$ with $L = T_{\mathrm{ini}} + N$ \cite{Elokda2021IJRNC,Huang2021TCST,Wang2022arXiv,wang2022experimental}. This choice typically leads to a very large value of $T$, making the optimization problem \eqref{equ_DeePC} large-scale and nontrivial to solve efficiently.
\end{remark}

\section{Minimum-dimension DeePC} \label{sec_svd}
In this section, we observe that the Hankel matrix~\eqref{equ_inputoutputHankel}~is always low-rank. This observation allows us to use a smaller data matrix to represent the input/output behavior of system \eqref{equ_LTI}, which can be viewed as a \textit{minimum-dimension} version of the fundamental lemma. We then introduce a procedure based on the singular-value decomposition (SVD) of the Hankel matrix~\eqref{equ_inputoutputHankel} to formulate a \textit{minimum-dimension} version~of~DeePC, addressing the dimension issues in both Remarks \ref{remark-1} and \ref{remark-2}.

\subsection{Minimum-Dimension Fundamental Lemma} 
The fundamental lemma plays an essential role in the standard DeePC~\eqref{equ_DeePC}. Indeed, each column of the Hankel matrix~\eqref{equ_inputoutputHankel} is a length-$L$ trajectory of system~\eqref{equ_LTI}, which can be regarded as a motion primitive. Lemma~\ref{lemma_Willems}~guarantees that any length-$L$ trajectory can be constructed via a linear combination of these motion primitives when the input sequence is persistently exciting. Note that the Hankel matrix 
$$
\mathcal{H}_{L} = \begin{bmatrix}
	\mathcal{H}_{L}(u^{\mathrm{d}}_{[0,T-1]}) \\
	\mathcal{H}_{L}(y^{\mathrm{d}}_{[0,T-1]})
\end{bmatrix} \in \mathbb{R}^{(m+p)L \times (T - L + 1)}
$$
is always low-rank, meaning that many motion primitives in the columns of~\eqref{equ_inputoutputHankel}  are redundant. We can thus use less, more representative motion primitives to represent the input-output behavior of system \eqref{equ_LTI}.   

\begin{lemma}\label{lemma_minimum}
	Consider a controllable LTI system \eqref{equ_LTI} and assume that the input sequence $u^{\mathrm{d}}_{[0,T-1]}$ is persistently exciting of order $n+L$. The following statements hold:
	\begin{enumerate}
		\item The rank of the Hankel matrix~\eqref{equ_inputoutputHankel} satisfies
		\begin{equation} \label{eq:rank-of-hankel}
			{
				r:= \mathrm{rank}\left(\begin{bmatrix}
					\mathcal{H}_{L}(u^{\mathrm{d}}_{[0,T-1]}) \\
					\mathcal{H}_{L}(y^{\mathrm{d}}_{[0,T-1]})
				\end{bmatrix}\right) \leq mL + n.
			}
		\end{equation}
		\item Suppose $\bar{\mathcal{H}}_{L} \in \mathbb{R}^{(m+p)L \times r}$ has the same range~space with the Hankel matrix~\eqref{equ_inputoutputHankel}. Then, any length-$L$ sequence $(u_{[0, L-1]}, y_{[0,L-1]})$ is an input/output trajectory of \eqref{equ_LTI} if and only if we have
		\begin{equation} \label{eq:fundamental-lemma-minimum}
			\begin{bmatrix}
				u_{[0,L-1]} \\ y_{[0,L-1]}
			\end{bmatrix} = \bar{\mathcal{H}}_{L} \bar{g}
		\end{equation}
		for some real vector $\bar{g} \in \mathbb{R}^{r}$. 
	\end{enumerate}
\end{lemma}

This result is known in the literature (see e.g., \cite[Appendix A]{Markovsky2021ARC}), but, to our knowledge, has not been utilized in reformulating DeePC to reduce the dimension for improved computational complexity. We give a brief proof of Lemma~\ref{lemma_minimum} below by observing the following relationship
\begin{equation} \label{equ_hankel_para}
	\begin{bmatrix}
		\mathcal{H}_{L}(u^{\mathrm{d}}_{[0,T-1]}) \\ \hline
		\mathcal{H}_{L}(y^{\mathrm{d}}_{[0,T-1]})
	\end{bmatrix} = 
	\begin{bmatrix}
		I_{mL} & 0_{mL\times n} \\ \hline
		\mathcal{T}_{L} & \mathcal{O}_{L}
	\end{bmatrix} 
	\begin{bmatrix}
		\mathcal{H}_{L}(u^{\mathrm{d}}_{[0,T-1]}) \\ \hline
		\mathcal{H}_{1}(x^{\mathrm{d}}_{[0,T-L]})
	\end{bmatrix},
\end{equation}
where the convolution matrix $\mathcal{T}_{L} \in \mathbb{R}^{pL \times mL}$ and the extended observability matrix $\mathcal{O}_{L} \in \mathbb{R}^{pL \times n}$ with $L$ block rows are
\[
{\small
	\begin{aligned}
		\mathcal{T}_L \!\!:=\!\! \begingroup\setlength\arraycolsep{2pt}\begin{bmatrix}
			D & 0 & 0 & \cdots & 0 \\
			CB & D & 0 & \cdots & 0 \\
			CAB & CB & D & \cdots & 0 \\
			\vdots & \vdots & \vdots & \ddots & \vdots \\
			CA^{L-2}B & CA^{L-3}B & CA^{L-4}B & \cdots & D
		\end{bmatrix}\!,
		\endgroup
		\mathcal{O}_L \!\!:=\!\! \begingroup\setlength\arraycolsep{2pt}\begin{bmatrix}
			C \\ 
			CA\\ 
			CA^2 \\
			\vdots \\ 
			CA^{L-1}
		\end{bmatrix}\!,
		\endgroup
	\end{aligned}
}
\]
and $\mathcal{H}_{1}(x^{\mathrm{d}}_{[0,T-L]}) = \begin{bmatrix} x^{\mathrm{d}}(0), \ldots,x^{\mathrm{d}}(T-L) \end{bmatrix}$. The factor 
$$
\begin{bmatrix}
	I_{mL} & 0_{mL\times n} \\
	\mathcal{T}_{L} & \mathcal{O}_{L}
\end{bmatrix}  \in \mathbb{R}^{(m+p)L \times (mL + n)}
$$
in \eqref{equ_hankel_para} has at most rank $mL + n$. We thus have the rank result~\eqref{eq:rank-of-hankel}. When $L$ is larger than the observability index (i.e., $\mathrm{rank}(\mathcal{O}_L) = n$), the rank result in~\eqref{eq:rank-of-hankel} can achieve the equality, i.e., $r = mL + n$. It is clear that the statement~\eqref{eq:fundamental-lemma-minimum} is equivalent to the statement~\eqref{eq:fundamental-lemma}. We note that the matrix $\bar{\mathcal{H}}_L$ in~\eqref{eq:fundamental-lemma-minimum} \textit{does not need to have a Hankel structure} as the matrix in~\eqref{equ_inputoutputHankel}, as long as they have the same range space. Finally, Lemma~\ref{lemma_minimum} can be extended to other matrix structures such as Page matrix \cite{Coulson2021TAC,Huang2021TCST} and mosaic-Hankel matrix \cite{Van2020CSL}; we present the details
in the appendix for completeness.  
\begin{remark}[Mini. dimension of the fundamental lemma] \label{remark:minimum}
	The Hankel matrix~\eqref{equ_inputoutputHankel} in Lemma~\ref{lemma_Willems} has $T-L+1$ motion primitives, while in Lemma~\ref{lemma_minimum}, the number of motion~primitives in $\bar{\mathcal{H}}_L$ has been reduced to $r \leq mL + n$. This upper bound is inherent to system dimensions and independent~of~the data length $T$. We can show from~\eqref{eq:g-dimension} that $T-L+1 -r \geq mn$, i.e., the dimension reduction is at least $mn$. It is clear that the column dimension of $\bar{\mathcal{H}}_L$ is minimum in order to guarantee the behavior representation of system~\eqref{equ_LTI} under the persistency excitation of the input $u^{\mathrm{d}}_{[0,T-1]}$. Thus, \eqref{eq:fundamental-lemma-minimum} can be viewed as a minimum-dimension version of Lemma~\ref{lemma_Willems}. 
\end{remark}

\begin{remark}[SVD-based dimension reduction] \label{remark:SVD}
	A standard way to generate $\bar{\mathcal{H}}_L\in \mathbb{R}^{(m+p)L \times r}$ is based on the SVD of the Hankel matrix~\eqref{equ_inputoutputHankel}. In particular, we let 
	\begin{equation} \label{eq:SVD}
		\begin{bmatrix}
			\mathcal{H}_{L}(u^{\mathrm{d}}_{[0,T-1]}) \\
			\mathcal{H}_{L}(y^{\mathrm{d}}_{[0,T-1]})
		\end{bmatrix} = \underbrace{\begin{bmatrix}
				W_1 & W_2
		\end{bmatrix}}_{W} \underbrace{\begin{bmatrix}
				\Sigma_1 & 0 \\ 0 & 0
		\end{bmatrix}}_{\Sigma} \underbrace{\begin{bmatrix}
				V_1 & V_2
			\end{bmatrix}^\tr}_{V^{\tr}},
	\end{equation}
	where $WW^\tr= W^{\tr}W = I_{(m+p)L}$ and $VV^\tr = V^{\tr}V = I_{T - L+1}$, and $\Sigma_1 \in \mathbb{R}^{r \times r}$ contains the top $r$ non-zero singular values. The choice
	\begin{equation} \label{eq:newHL}
		{
			\bar{\mathcal{H}}_{L} = \mathcal{H}_{L}V_{1} = W_1 \Sigma_{1} \in \mathbb{R}^{(m+p)L \times r}
		}
	\end{equation}
	satisfies the range space condition in Lemma~\ref{lemma_minimum}.   
\end{remark}

\subsection{Minimum-Dimension DeePC}
After collecting the input/output data sequences $u^{\mathrm{d}}_{[0,T-1]}$,  $y^{\mathrm{d}}_{[0,T-1]}$ and forming the Hankel data matrices in \eqref{eq:Hankel}, we can apply the SVD technique in Remark \ref{remark:SVD}  to compute a new data library $\bar{\mathcal{H}}_L\in \mathbb{R}^{(m+p)L \times r}$ in \eqref{eq:newHL} and consequently achieve dimension reduction in DeePC.    

In particular, we replace the DeePC problem in \eqref{equ_DeePC} by
\begin{align}\label{equ_DeePC_new}
	\min_{\bar{g},u,y,\sigma_{u},\sigma_{y}} & \left\|y - y_{r}\right\|_{Q}^{2} + \left\|u\right\|_{R}^{2} 
	+ \lambda_{u}\left\|\sigma_{u} \right\|_{2}^{2} \nonumber \\
	&\qquad \qquad \qquad \qquad  + \lambda_{y}\left\|\sigma_{y} \right\|_{2}^{2} + \lambda_{g}\left\|\bar{g} \right\|_{2}^{2}  \nonumber
	\\
	\mathrm{subject~to} &\quad  \bar{\mathcal{H}}_{L} \bar{g} = \begin{bmatrix}
		u_{\mathrm{ini}} \\ u \\ y_{\mathrm{ini}} \\ y
	\end{bmatrix} + \begin{bmatrix}
		\sigma_{u} \\ 0 \\ \sigma_{y} \\ 0
	\end{bmatrix},
	u \in\mathcal{U}, y \in\mathcal{Y}. 
\end{align}
We refer to~\eqref{equ_DeePC_new} as the \textit{minimum-dimension} version of DeePC, since the column dimension of $\bar{\mathcal{H}}_L$ is minimum to guarantee the behavior representation of LTI systems. The optimization variable $g$ in~\eqref{equ_DeePC} has a dimension of $T-L+1$, while in~\eqref{equ_DeePC_new}, the dimension of the optimization variable $\bar{g}$ has been reduced to $r$. As discussed in Remark~\ref{remark:minimum}, the dimension reduction holds a lower bound of $mn$, which can be significant when the system has a large internal system state dimension $n$ or input dimension $m$. In addition, the dimension reduction scheme assumes a prominent role in practical applications (e.g.,~\cite{Elokda2021IJRNC,Huang2021TCST,Wang2022arXiv,wang2022experimental}), especially when the value of $n$ and the bound on $T_{\mathrm{ini}}$ are unknown and $T$ needs to be sufficiently large as discussed in Remark~\ref{remark-2}.

It is not difficult to see that the objective functions in~\eqref{equ_DeePC} and \eqref{equ_DeePC_new} are strongly convex when $\lambda_{u}>0, \lambda_{y}>0,\lambda_{g}>0$. Both \eqref{equ_DeePC} and \eqref{equ_DeePC_new} have a unique optimal solution if $u \in\mathcal{U}, y \in\mathcal{Y}$ are convex~constraints. Indeed, we can show the equivalence of the unique optimal solution $(u^*, y^*, \sigma_{u}^*, \sigma_{y}^*)$ for \eqref{equ_DeePC} and \eqref{equ_DeePC_new}.

\begin{theorem} \label{theorem1}
	Suppose $\bar{\mathcal{H}}_{L}$ is generated with \eqref{eq:newHL} that shares the same range space with $\mathcal{H}_{L}$, the parameters $\lambda_{u}$, $\lambda_{y}$, and $\lambda_{g}$ are positive, and $\mathcal{U}$ and $\mathcal{Y}$ are convex polytopes. If $g^{*}$ minimizes \eqref{equ_DeePC}, then $\bar{g}^{*} = V_{1}^{\tr}g^{*}$ is the minimizer of \eqref{equ_DeePC_new}. Moreover, \eqref{equ_DeePC} and \eqref{equ_DeePC_new} have the same optimal solution $(u^*, y^*, \sigma_{u}^*, \sigma_{y}^*)$.
\end{theorem}

The proof details are postponed to the appendix. The idea is to utilize the KKT condition of \eqref{equ_DeePC} and $\eqref{equ_DeePC_new}$ to establish the connection between $g^{*}$ and $\bar{g}^{*}$. We note that the polyhedral constraints $\mathcal{U}$ and $\mathcal{Y}$  allow simple KKT conditions. Then, leveraging matrix properties obtained through SVD in \eqref{eq:SVD}, we prove the equivalence of the optimal solution $(u^*, y^*, \sigma_{u}^*, \sigma_{y}^*)$ between \eqref{equ_DeePC} and \eqref{equ_DeePC_new}.

\begin{remark}[Cases beyond deterministic LTI systems]
	One main motivation of the regularization in~\eqref{equ_DeePC} is to handle systems beyond deterministic LTI with measurement and process noises \cite[Section VI]{Coulson2019ECC} (see \cite{Markovsky2021ARC,Coulson2021TAC} for more details). In this case, the Hankel matrix \eqref{equ_inputoutputHankel} is typically full rank. However, it is unnecessary and not beneficial to use the full range space of the Hankel matrix in \eqref{equ_DeePC} (which is one main reason for using regularization on $\sigma_u$, $\sigma_y$, and $g$ to select important columns). 
	In this case, we can still use SVD-based techniques to extract the dominant range space of the Hankel matrix~\eqref{equ_inputoutputHankel}. In particular, after performing SVD~\eqref{eq:SVD}, we plot singular values in $\Sigma$ by using a base-10 logarithmic scale. Via the distribution of singular values in descending order, an appropriate selection of $r$ can be informed by the turning point, which indicates the transition from singular values/vectors that represent principal patterns to those that are insignificant. This leads to a much condensed data library $\bar{\mathcal{H}}_L \in \mathbb{R}^{(m+p)L \times r}$ with a reduced column number. This strategy is generally robust to the length $T$ of the input sequence  $u^{\mathrm{d}}_{[0,T-1]}$. In practice, the length $T$ can be chosen as large as possible, but the value of $r$ via SVD may still remain small. This addresses the dimension issue in Remark~\ref{remark-2}. 
\end{remark}  

\section{Numerical experiments} \label{sec_simulation}
This section presents two numerical examples to demonstrate the effectiveness of our SVD-based approach. All computations are performed in MATLAB 2022a on a laptop with an Intel i7-10710U CPU with 6 cores, 1.6 GHz clock rate and 16 GB RAM. The DeePC problem can be transformed into the quadratic problem and is solved with the MATLAB function \texttt{quadprog}.

\subsection{Linear System Case Study}
We first consider an LTI system with matrices $(A, B$, $C, D)$ as follows
\[
\begingroup 
\setlength\arraycolsep{2pt}
\begin{aligned}
	A &= \begin{bmatrix}
		0.921 & 0 & 0.041 & 0 \\
		0 & 0.918 & 0 & 0.033 \\
		0 & 0 & 0.924 & 0 \\
		0 & 0 & 0 & 0.937
	\end{bmatrix},  B = \begin{bmatrix}
		0.017 & 0.001 \\
		0.001 & 0.023 \\
		0 & 0.061 \\
		0.072 & 0
	\end{bmatrix},
	\\
	C &= \begin{bmatrix}
		1 & 0 & 0 & 0 \\
		0 & 1 & 0 & 0
	\end{bmatrix}, D = 0_{2\times 2}.
\end{aligned}
\endgroup
\]
The control objective is to track the setpoint $y_r(t) = \begin{bmatrix}
	0.65, 0.77
\end{bmatrix}^{\tr}$, and the input/output constraints are
${\small\begin{bmatrix}
		-2, -2
	\end{bmatrix}^{\tr}} \le u(t) \le \begin{bmatrix}
	2, 2
\end{bmatrix}^{\tr}$ and ${\small\begin{bmatrix}
		-2, -2
	\end{bmatrix}^{\tr}\le y(t) \le \begin{bmatrix}
		2, 2
	\end{bmatrix}^{\tr}}$, respectively. 

\begin{figure}[!t]
	\centering
	\includegraphics[width=0.32\textwidth]{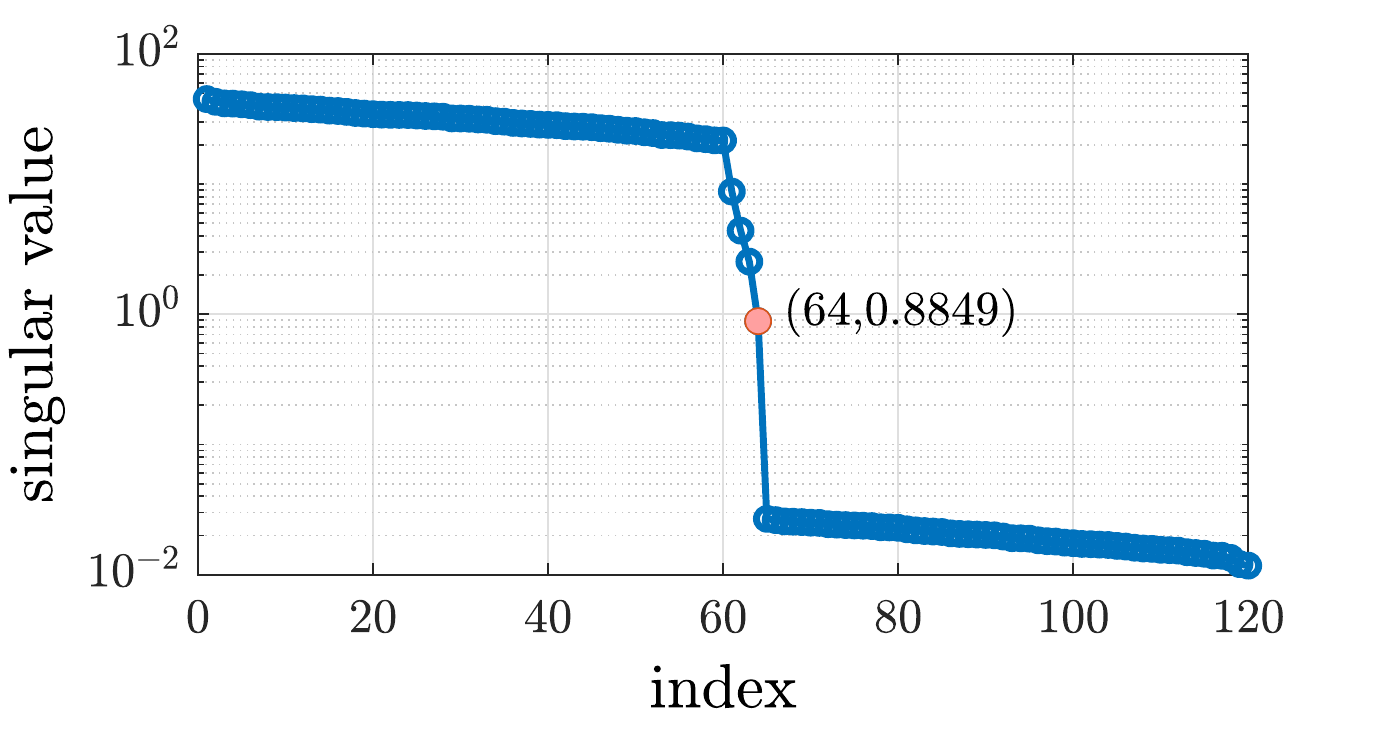}
	\vspace{-3mm}
	\caption{Singular value distribution of $\mathcal{H}_{L}$ in linear system case.}
	\label{fig_svd_linearCase}
	\vspace{-2mm}
\end{figure}

In an offline experiment, an input/output trajectory of length $T=400$ is collected, where the input $u(t)$ is chosen randomly from $\begin{bmatrix}
	-3, 3
\end{bmatrix}^{\tr}$, and the output $y(t)$ are subject to uniformly distributed noise over $\begin{bmatrix}
	-0.002, 0.002
\end{bmatrix}^{\tr}$. The time horizons for the previous data sequence and future data sequence are chosen as $T_{\mathrm{ini}} = 10$ and $N = 20$, and thus $L=T_{\mathrm{ini}} + N=30$. The weighting matrices $Q$, $R$ and the regularization parameters $\lambda_{u}$, $\lambda_{y}$, $\lambda_{g}$ are chosen as $Q = 35\cdot I_{40}$, $R = 10^{-4}I_{40}$, $\lambda_{u} = 10^{6}$, $\lambda_{y} = 10^{4}$, and $\lambda_{g} = 10^{2}$, respectively. We can calculate that the Hankel matrix $\mathcal{H}_{L}$ is of size $120\times 371$, while its rank should be at most $mL+n=64$ under the noise-free deterministic case. 
The singular value distribution of $\mathcal{H}_{L}$ (i.e., $\sigma_{i}$, $i=1, \cdots, 120$) is shown in Fig.~\ref{fig_svd_linearCase}. It is clear that the turning point is $64$, which is consistent with the analysis.

\begin{figure}[!t]
	\setlength{\abovecaptionskip}{0pt}
	\centering
	\subfigure[]{\label{fig_outputInput_H1_linearCase}
		\includegraphics[width=0.162\textwidth]{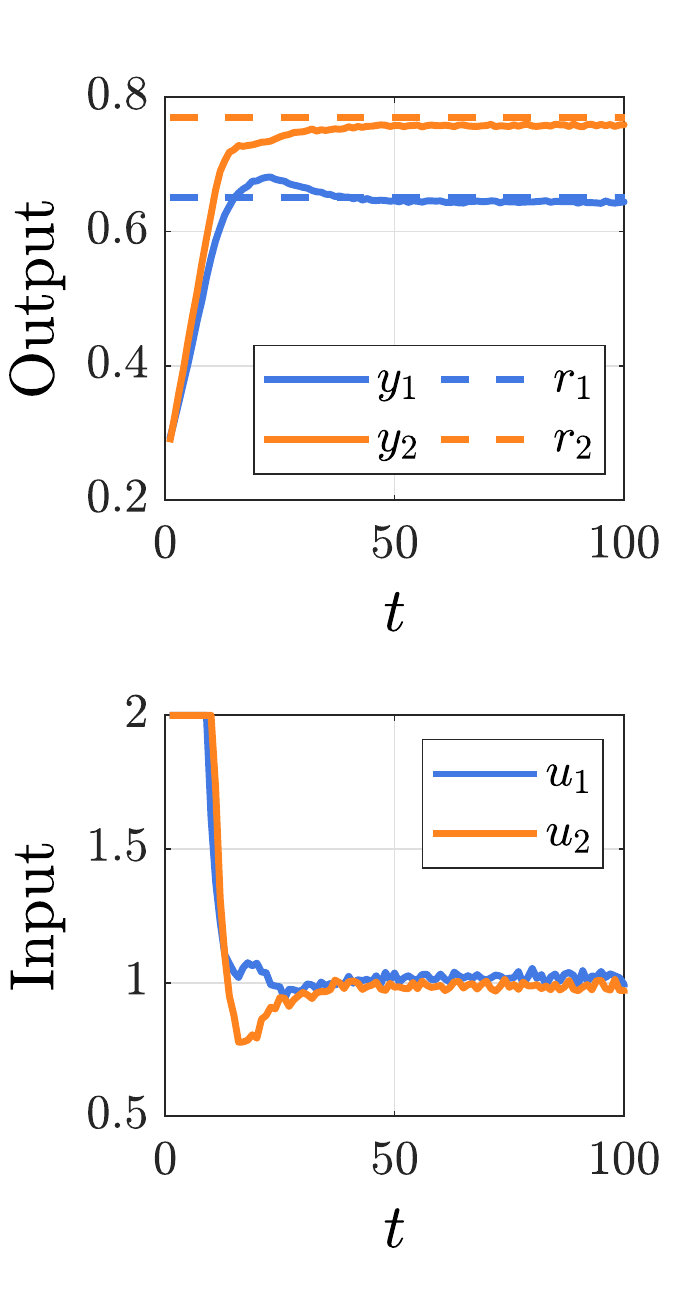}
	}
	\hspace{-6mm}
	\subfigure[]{\label{fig_outputInput_H2_linearCase}
		\includegraphics[width=0.162\textwidth]{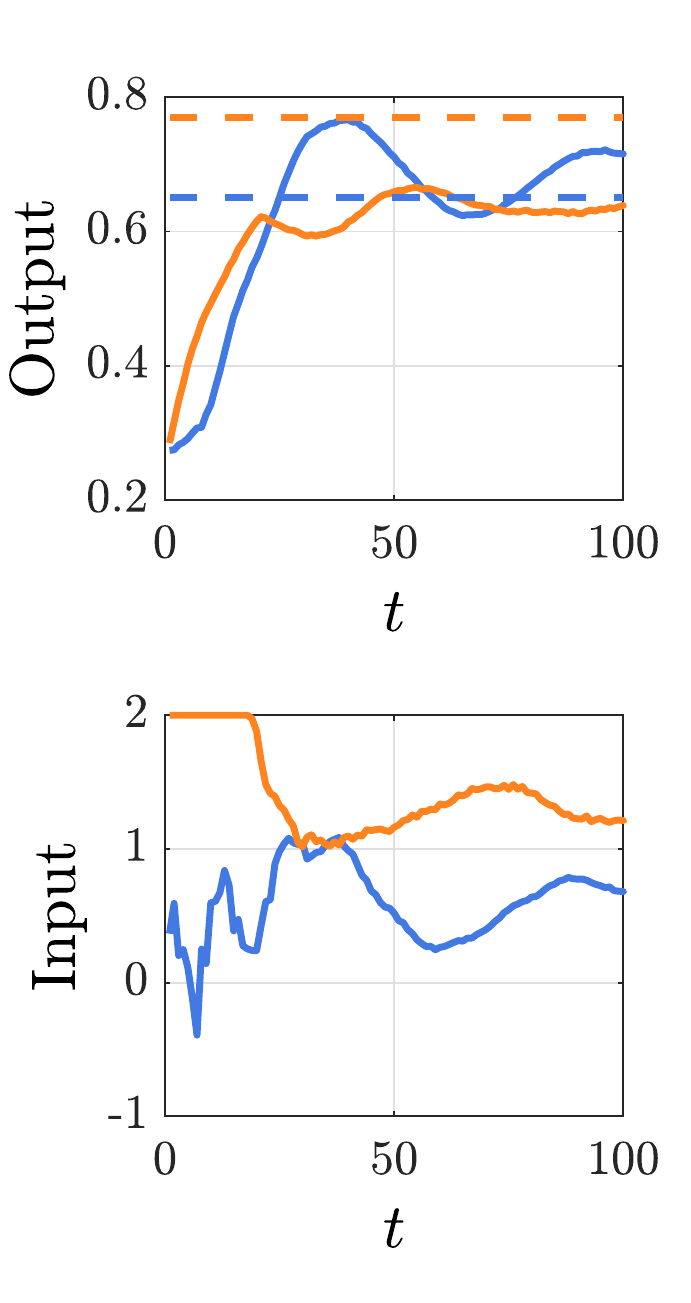}
	}
	\hspace{-6mm}
	\subfigure[]{\label{fig_outputInput_H3_linearCase}
		\includegraphics[width=0.162\textwidth]{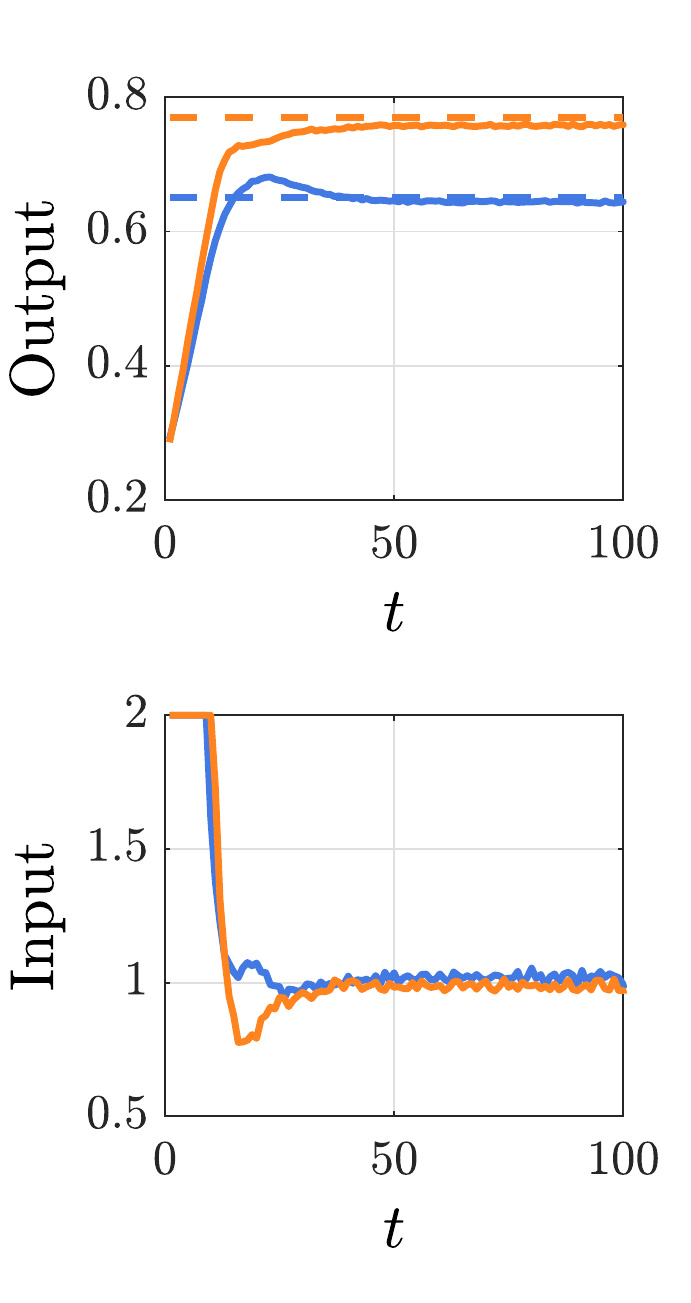}
	}
	\caption{Output and input evolution of the linear system, resulting from the application of the DeePC algorithm with (a) original data library $\mathcal{H}_{L}$, (b) truncation matrix $\mathcal{H}_{L,[1:r]}$, and (c) new data library $\bar{\mathcal{H}}_{L}$.}
	\label{fig_outputInput_linearCase}
	\vspace{-2mm}
\end{figure}

We extract the first $r=64$ columns and singular values from $W$ and $\Sigma$, respectively, to construct $\bar{\mathcal{H}}_{L}$ as discussed in Remark~\ref{remark:SVD}. For comparison, the original data library $\mathcal{H}_{L}$, its direct truncation matrix $\mathcal{H}_{L,[1:r]}$ (i.e., the first $r$ columns of $\mathcal{H}_{L}$ are extracted to construct the truncation matrix), and the new data library $\bar{\mathcal{H}}_{L}$ are used in DeePC. Fig.~\ref{fig_outputInput_linearCase} shows the simulation results. It is clear that both the original data library $\mathcal{H}_{L}$ and the low-dimensional library $\bar{\mathcal{H}}_{L}$ can be well embedded into the DeePC framework to achieve satisfactory control performance, while the truncation matrix $\mathcal{H}_{L,[1:r]}$ fails to regulate the output signals. 

In Table~\ref{table_comparison}, we further list the accumulative cost (i.e., $\sum \left( \left\|y - y_r\right\|_{Q}^{2}+\left\|u\right\|_{R}^{2}\right)$) and the average computation time per iteration of the DeePC algorithm based on $\mathcal{H}_{L}$ and $\bar{\mathcal{H}}_{L}$. 
Since $\bar{\mathcal{H}}_{L}$ has a much lower column dimension than $\mathcal{H}_{L}$ (64 vs. 371), its DeePC implementation admits more efficient computation (17 ms vs. 48 ms), while maintaining similar closed-loop performance thanks to the SVD procedure that extracts principle patterns.

\begin{table}[!t]
	\begin{center}
		\setlength{\abovecaptionskip}{3pt}
		\caption{Comparison in the Linear System Case Study}\label{table_comparison}
		\scalebox{0.95}{
			\begin{threeparttable}
				\setlength{\tabcolsep}{2mm}{
					\begin{tabular}{c c c}
						\hline
						\hline
						& DeePC & Minimum-Dimension DeePC \\
						\hline
						Dimension of the variable $g$/$\bar{g}$ & 371 & 64
						\\
						Average Computation Time [ms] & 48.34 & 17.34 \\
						Accumulative Cost & 169.63 & 169.64 \\
						\hline
						\hline
				\end{tabular}}
			\end{threeparttable}
		}
	\end{center}
	\vspace{-4mm}
\end{table}

\subsection{Case Study with Nonlinear Traffic System}
To evaluate the performance of the proposed minimum-version DeePC scheme for nonlinear systems, we consider the leading cruise control (LCC) of connected and autonomous vehicles (CAVs) in mixed traffic scenarios \cite{Wang2022arXiv}.

\begin{figure}[!t]
	\centering
	\setlength{\abovecaptionskip}{2pt}
	\includegraphics[scale=0.46]{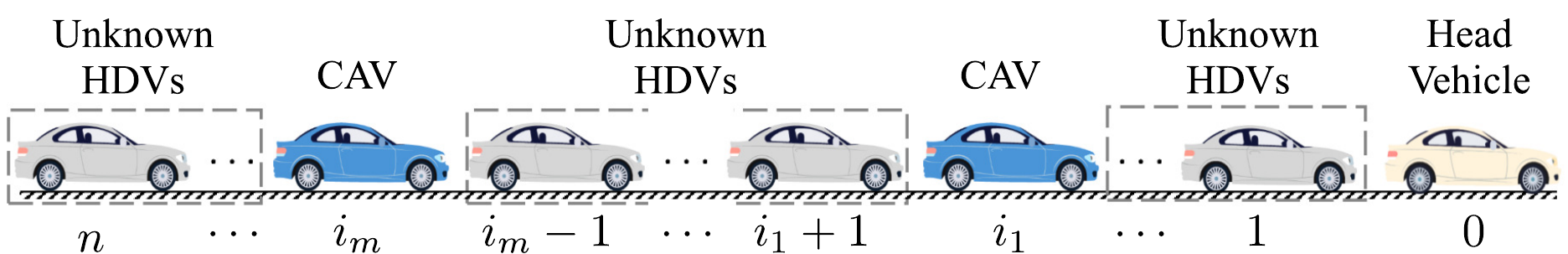}
	\caption{Schematic of mixed traffic system with $n+1$ vehicles.}
	\label{fig_LCC}
	\vspace{-2mm}
\end{figure}

As shown in Fig.~\ref{fig_LCC}, the mixed traffic consists of $n+1$ vehicles: one head vehicle (indexed as $0$), $m$ CAVs, and $n-m$ human-driven vehicles (HDVs). Let ${\small \Omega=\{1,2,\ldots,n\}}$ be the set of vehicle indices ordered from front to end. The sets of CAV indices and HDV indices are denoted by ${\small \Omega_{C}=\{i_1,\ldots,i_m\}\subseteq \Omega}$ and ${\small \Omega_{H}=\{j_1,\ldots,j_{n-m}\}= \Omega \backslash \Omega_{C}}$, respectively, where ${\small i_1 < \ldots < i_m}$ and ${\small j_1 < \ldots < j_{n-m}}$. The spacing error, velocity error and acceleration of the $i$-th vehicle at time $t$ is denoted as $\tilde{s}_i(t)$, $\tilde{v}_i(t)$ and $u_i(t)$, respectively. The system state, input, and output of the mixed traffic are given by, respectively,  $x(t) = \begin{bmatrix}
	\tilde{s}_{1}(t), \tilde{v}_{1}(t), \cdots, \tilde{s}_{n}(t), \tilde{v}_{n}(t)
\end{bmatrix}^{\tr}$, and
$$
\begin{aligned}
	u(t) &= \begin{bmatrix}
		u_{i_1}(t), \cdots, u_{i_m}(t)
	\end{bmatrix}^{\tr},\\
	y(t) &= \begin{bmatrix}
		\tilde{v}_{1}(t), \cdots, \tilde{v}_{n}(t), \tilde{s}_{i_1}(t), \cdots, \tilde{s}_{i_m}(t) \end{bmatrix}^{\tr}.
\end{aligned}
$$
A method called the DeeP-LCC algorithm has been developed in \cite{Wang2022arXiv} to achieve safe and optimal control of CAVs. 

We test our developed approach under three fleet structures with different numbers of CAVs and HDVs. These three fleets are described by
\[
{\small
	\begingroup 
	\setlength\arraycolsep{2pt}
	\begin{aligned}
		&\text{Fleet 1:}\; n=3, m=1, \Omega_{C}=\left\{ 2 \right\}, \Omega_{H}=\left\{1, 3 \right\},
		\\
		&\text{Fleet 2:}\; n=5, m=2, \Omega_{C}=\left\{ 2, 4 \right\}, \Omega_{H}=\left\{1, 3, 5 \right\},
		\\
		&\text{Fleet 3:}\; n=8, m=2,
		\Omega_{C}=\left\{ 3, 6 \right\}, \Omega_{H}=\left\{1, 2, 4, 5, 7, 8 \right\}.
	\end{aligned}
	\endgroup
}
\]
As suggested in \cite{Wang2022arXiv}, the length for the pre-collected data is chosen as $T=2000$, $T=3000$, and $T=4000$,~respectively, under these three fleet structures. The remaining parameters used to set up the DeePC are selected as the same with~\cite{Wang2022arXiv}. A safety-critical scenario is considered in our simulation; see \cite[Section VI-C]{Wang2022arXiv} for a detailed description.

\begin{figure}[!t]
	\centering
	\includegraphics[width=0.4\textwidth]{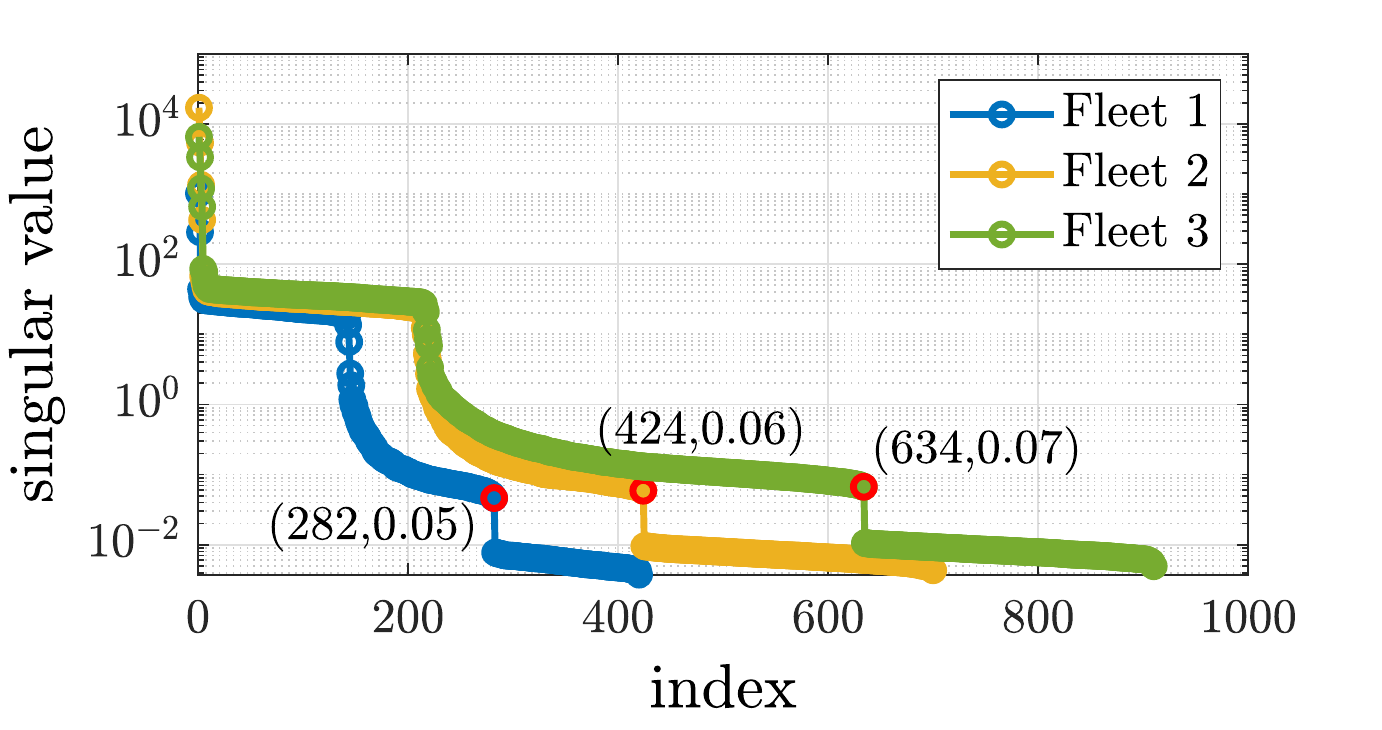}
	\vspace{-3mm}
	\caption{Singular value distribution of $\mathcal{H}_{L}$ for Fleet~1, Fleet~2, and Fleet~3.}
	\label{fig_svd_lcc}
	\vspace{-2mm}
\end{figure}

For these three fleet structures, $\mathcal{H}_{L}$ has a dimension of $420\times 1931$, $700\times 2931$, and $910\times 3931$, respectively, and its corresponding singular values distribution is shown in Fig.~\ref{fig_svd_lcc}. It is clear that the data Hankel matrices have a low rank, and accordingly, we select $r$ as $r=282$, $r=424$, and $r=634$, respectively. Under each fleet, the data library $\mathcal{H}_{L}$ and the low-dimensional data library $\bar{\mathcal{H}}_{L}$ are utilized in DeePC for the control of CAVs. We use fuel consumption, average absolute velocity error (AAVE) \cite{Wang2022arXiv}, and average computation time per iteration to depict the performance. 

The results are summarized in Table~\ref{table_comparison_lcc}. It is noted that under different fleet structures, the proposed minimum-dimension DeePC consistently offers an order of magnitude faster computational time while achieving similar control performance as compared to the original DeePC. This confirms that, in the context of nonlinear systems, the SVD-based approach can successfully extract the principle patterns from $\mathcal{H}_{L}$, and thus the new data library $\bar{\mathcal{H}}_{L}$ preserves the critical information of $\mathcal{H}_{L}$, while allowing for a more efficient representation of the system behavior. Therefore, the resulting minimum-dimension DeePC can accomplish better computational efficiency without compromising the control performance.

\begin{table}[!t]
	\begin{center}
		\caption{Comparison in the Nonlinear System Case}\label{table_comparison_lcc}
		\scalebox{0.92}{
			\begin{threeparttable}
				\setlength{\tabcolsep}{1mm}{
					\begin{tabular}{c c c || c c c}
						\hline
						\hline
						& DeePC & \begin{tabular}{@{}c@{}}Minimum-Dimension \\ DeePC\end{tabular} & & DeePC &  \begin{tabular}{@{}c@{}}Minimum-Dimension \\ DeePC\end{tabular}
						\\
						\hline
						\multicolumn{3}{l||}{{\textbf{Dimension of} $g$/$\bar{g}$}} & \multicolumn{3}{l}{{\textbf{Average Computation Time} [s]}}
						\\
						Fleet 1 & 1931 & 282 & Fleet 1 & 1.02 & 0.06 \\
						Fleet 2 & 2931 & 424 & Fleet 2 & 3.59 & 0.25 \\
						Fleet 3 & 3931 & 634 & Fleet 3 & 7.41 & 0.36 \\
						\hline
						\multicolumn{3}{l||}{{\textbf{Fuel Consumption} [mL]}} & \multicolumn{3}{l}{{\textbf{AAVE} [$10^{-2}$]}}
						\\
						Fleet 1 & 94.83 & 94.79 & Fleet 1 & 10.77 & 10.78 \\
						Fleet 2 & 179.19 & 179.10 & Fleet 2 & 14.32 & 14.31 \\
						Fleet 3 & 303.94 & 304.06 & Fleet 3 & 21.39 & 21.40 \\
						\hline
						\hline
				\end{tabular}}
			\end{threeparttable}
		}
	\end{center}
	\vspace{-6mm}
\end{table}

\section{Conclusions} \label{sec_conclusion}
This letter has presented an SVD-based strategy to reduce the dimension of the optimization problem in DeePC. It is known that a large data library measured from an LTI system can be refined into a low-dimensional one for the non-parametric representation of system behavior. Based on this observation, we have proposed an SVD-based approach to extract the main components from the large data library and subsequently reduce the optimization problem dimension in the DeePC formulation. Simulation results showcase that by using the proposed method, the computation efficiency of the DeePC algorithm can be an order of magnitude faster without sacrificing the control performance.
				
\input{appendix}
				
				
\bibliographystyle{IEEEtran}
\bibliography{IEEEabrv,reference}
\balance
				
\end{document}

%% file: appendix.tex
\appendices

\section{ Minimum-Dimension Non-Parametric Representation With Page Matrix}
Page matrix is another effective structure to store raw data. We now introduce its definition and persistently exciting condition.
The Page matrix of depth $k$ associated with $\omega_{\left[i, j\right]}$ is defined as
\[\begingroup 
\setlength\arraycolsep{2pt}
\begin{aligned}
	&\mathcal{P}_{k}(\omega_{\left[i, j\right]}):= 
	\\
	&\begin{bmatrix}
		\omega(i) \!&\! \omega(i+k) \!&\! \cdots \!&\! \omega(i+\lfloor \frac{j-i+1}{k}-1 \rfloor k) \\
		\omega(i+1) \!&\! \omega(i+k+1) \!&\! \cdots \!&\! \omega(i+\lfloor \frac{j-i+1}{k}-1 \rfloor k+1)\\
		\vdots \!&\! \vdots \!&\! \ddots \!&\! \vdots \\
		\omega(i+k-1) \!&\! \omega(i+2k-1) \!&\! \cdots \!&\! \omega(i+\lfloor \frac{j-i+1}{k} \rfloor k-1)
	\end{bmatrix},
\end{aligned}	\endgroup
\]
where $\lfloor \cdot \rfloor$ is the floor function which rounds its argument down to the nearest integer.
\begin{definition} [{\!\cite{Coulson2021TAC}}]
	The sequence $\omega_{\left[i, j\right]}$ is said to be $k$-Page exciting of order $l$, where $l\in \mathbb{Z}$, if the matrix
	\[
	\begin{bmatrix}
		\mathcal{P}_{k}(\omega_{\left[i, j-(l-1)k\right]})
		\\
		\mathcal{P}_{k}(\omega_{\left[k+i, j-(l-2)k\right]})
		\\
		\vdots
		\\
		\mathcal{P}_{k}(\omega_{\left[k(l-1)+i, j\right]})
	\end{bmatrix}
	\] 
	has full row rank.
\end{definition}

Given the input/output trajectory $(u^{\mathrm{d}}_{[0,T-1]}, y^{\mathrm{d}}_{[0,T-1]})$ of length $T$, the Page matrices $\mathcal{P}_{L}(u^{\mathrm{d}}_{[0,T-1]})$ and $\mathcal{P}_{L}(y^{\mathrm{d}}_{[0,T-1]})$ are given by
\begin{equation} \label{equ_inputoutputPage}
	\begingroup 
	\setlength\arraycolsep{2pt} \small 
	\begin{bmatrix}
		\mathcal{P}_{L}(u^{\mathrm{d}}_{[0,T-1]}) \\
		\mathcal{P}_{L}(y^{\mathrm{d}}_{[0,T-1]})
	\end{bmatrix} \!:= \! 
	\begin{bmatrix} u^{\mathrm{d}}(0) & u^{\mathrm{d}}(L) & \cdots & u^{\mathrm{d}}(\lfloor \frac{T}{L} - 1 \rfloor L) \\
		\vdots & \vdots & & \vdots \\
		u^{\mathrm{d}}(L\!-\!1) & u^{\mathrm{d}}(2L\!-\!1) & \cdots & u^{\mathrm{d}}(\lfloor \frac{T}{L} \rfloor L-1) \\
		y^{\mathrm{d}}(0) & y^{\mathrm{d}}(L) & \cdots & y^{\mathrm{d}}(\lfloor \frac{T}{L} - 1 \rfloor L) \\
		\vdots & \vdots & & \vdots \\
		y^{\mathrm{d}}(L\!-\!1) & y^{\mathrm{d}}(2L\!-\!1) & \cdots & y^{\mathrm{d}}(\lfloor \frac{T}{L} \rfloor L-1)  
	\end{bmatrix}\!.
	\endgroup
\end{equation}

\begin{lemma}[\cite{Coulson2021TAC}]\label{lemma_Page}
	Consider a controllable LTI system \eqref{equ_LTI} and assume that the input sequence $u^{\mathrm{d}}_{[0,T-1]}$ is $L$-Page exciting of order $n+1$. Then, any length-$L$  $(u_{[0, L-1]}, y_{[0,L-1]})$ is an input/output trajectory of \eqref{equ_LTI} if and only if we have
	\begin{equation} \label{eq:fundamental-lemma-page}
		\begin{bmatrix}
			u_{[0,L-1]} \\ y_{[0,L-1]}
		\end{bmatrix} = \begin{bmatrix}
			\mathcal{P}_{L}(u^{\mathrm{d}}_{[0,T-1]}) \\
			\mathcal{P}_{L}(y^{\mathrm{d}}_{[0,T-1]})
		\end{bmatrix} g
	\end{equation}
	for some real vector $g \in \mathbb{R}^{\lfloor \frac{T}{L} \rfloor}$. 
\end{lemma}

For the input sequence $u^{\mathrm{d}}_{[0,T-1]}$ to be $L$-Page exciting of order $n+1$, one requires that $T\ge L((mL+1)(n+1)-1)$. Thus, the Page matrix 
\[
\mathcal{P}_{L} = \begin{bmatrix}
	\mathcal{P}_{L}(u^{\mathrm{d}}_{[0,T-1]}) \\
	\mathcal{P}_{L}(y^{\mathrm{d}}_{[0,T-1]})
\end{bmatrix} \in \mathbb{R}^{(m+p)L \times (\lfloor \frac{T}{L} \rfloor L - L + 2)}
\]
is always low rank, and less motion primitives can be used for the non-parametric representation of system \eqref{equ_LTI}.

\begin{lemma}\label{lemma_Page_minimum}
	Consider a controllable LTI system \eqref{equ_LTI} and assume that the input sequence $u^{\mathrm{d}}_{[0,T-1]}$ is $L$-Page exciting of order $n+1$. 
	The following statements hold:
	\begin{enumerate}
		\item The rank of the Page matrix~\eqref{equ_inputoutputPage} satisfies
		\begin{equation} \label{eq:rank-of-page}
			r:= \mathrm{rank}\left(\begin{bmatrix}
				\mathcal{P}_{L}(u^{\mathrm{d}}_{[0,T-1]}) \\
				\mathcal{P}_{L}(y^{\mathrm{d}}_{[0,T-1]})
			\end{bmatrix}\right) \leq mL + n.
		\end{equation}
		\item Suppose $\bar{\mathcal{P}}_{L} \in \mathbb{R}^{(m+p)L \times r}$ has the same range~space with the Page matrix~\eqref{equ_inputoutputPage}. Then, any length-$L$ sequence $(u_{[0, L-1]}, y_{[0,L-1]})$ is an input/output trajectory of \eqref{equ_LTI} if and only if we have
		\begin{equation} \label{eq:page-lemma-minimum}
			\begin{bmatrix}
				u_{[0,L-1]} \\ y_{[0,L-1]}
			\end{bmatrix} = \bar{\mathcal{P}}_{L} \bar{g}
		\end{equation}
		for some real vector $\bar{g} \in \mathbb{R}^{r}$. 
	\end{enumerate}
\end{lemma}

By the system dynamics we have that
\begin{equation*}
	\begin{bmatrix}
		\mathcal{P}_{L}(u^{\mathrm{d}}_{[0,T-1]}) \\ \hline
		\mathcal{P}_{L}(y^{\mathrm{d}}_{[0,T-1]})
	\end{bmatrix} = 
	\begin{bmatrix}
		I_{mL} & 0_{mL\times n} \\ \hline
		\mathcal{T}_{L} & \mathcal{O}_{L}
	\end{bmatrix} 
	\begin{bmatrix}
		\mathcal{P}_{L}(u^{\mathrm{d}}_{[0,T-1]}) \\ \hline
		x^{\mathrm{d}}_{[0,L,\lfloor \frac{T}{L} \rfloor-1]}
	\end{bmatrix},
\end{equation*}
where $x^{\mathrm{d}}_{[0,L,\lfloor \frac{T}{L} \rfloor-1]} = \begin{bmatrix}
	x^{\mathrm{d}}(0), x^{\mathrm{d}}(L), \cdots, x^{\mathrm{d}}(\lfloor \frac{T}{L} \rfloor L-L)
\end{bmatrix}$. The proof of Theorem 2.1 in \cite{Coulson2021TAC} shows that the controllability assumption and the condition that $u^{d}_{[0,T-1]}$ is $L$-Page exciting of order $n+1$ imply
\[
\mathrm{rank}\left( \begin{bmatrix}
	\mathcal{P}_{L}(u^{\mathrm{d}}_{[0,T-1]}) \\ \hline
	x^{\mathrm{d}}_{[0,L,\lfloor \frac{T}{L} \rfloor-1]}
\end{bmatrix} \right) = mL+n.
\]
Since the matrix $$
\begin{bmatrix}
	I_{mL} & 0_{mL\times n} \\
	\mathcal{T}_{L} & \mathcal{O}_{L}
\end{bmatrix}  \in \mathbb{R}^{(m+p)L \times (mL + n)}
$$
has at most rank $mL + n$, we have the rank result \eqref{eq:rank-of-page}. Moreover, it is clear that the statement~\eqref{eq:page-lemma-minimum} is equivalent to the statement~\eqref{eq:fundamental-lemma-page} provided that $\bar{\mathcal{P}}_{L}$ has the same range~space with the Page matrix $\mathcal{P}_{L}$.

\section{ Minimum-Dimension Non-Parametric Representation With Mosaic-Hankel Matrix}
Both Hankel and Page matrices are used to arrange a single data sequence. When multiple system trajectories are given, mosaic-Hankel matrix can be applied to achieve non-parametric system representation. Let $(u^{\mathrm{d},i}_{[0,T_{i}-1]}, y^{\mathrm{d},i}_{[0,T_{i}-1]})$ be an input/output trajectory of \eqref{equ_LTI} for $i=1,2,\cdots, q$, where $q$ is the number of data sets. The mosaic-Hankel matrix of depth $k$ is defined as 
\begin{equation} \label{equ_mosaic}
	\begingroup 
	\setlength\arraycolsep{2pt} \small 
	\begin{bmatrix}
		\mathcal{H}_{k}(u^{\mathrm{d},1}_{[0,T_{1}-1]}) & \mathcal{H}_{k}(u^{\mathrm{d},2}_{[0,T_{2}-1]}) & \cdots & \mathcal{H}_{k}(u^{\mathrm{d},q}_{[0,T_{q}-1]})
	\end{bmatrix}.
	\endgroup
\end{equation}

\begin{definition} [{\cite{Van2020CSL}}]
	The input sequences  $u^{\mathrm{d},i}_{[0,T_{i}-1]}$ for $i=1,2,\cdots,q$ are called collectively persistently exciting of order $k$ if the mosaic-Hankel matrix \eqref{equ_mosaic} has full row rank.
\end{definition}

Based on the notion of collective persistency of excitation, the fundamental lemma can be extended to the case of multiple data sets.

\begin{lemma}[\cite{Van2020CSL}]\label{lemma_mosaic}
	Consider a controllable LTI system \eqref{equ_LTI} and assume that the input sequences $u^{\mathrm{d},i}_{[0,T_{i}-1]}$ are collectively persistently exciting of order $n+L$. Then, any length-$L$  $(u_{[0, L-1]}, y_{[0,L-1]})$ is an input/output trajectory of \eqref{equ_LTI} if and only if we have
	\begin{equation} \label{eq:fundamental-lemma-mosaic}
		\begin{bmatrix}
			u_{[0,L-1]} \\ y_{[0,L-1]}
		\end{bmatrix} = \begin{bmatrix}
			\mathcal{H}_{L}(u^{\mathrm{d},1}_{[0,T_{1}-1]}) & \cdots & \mathcal{H}_{L}(u^{\mathrm{d},q}_{[0,T_{q}-1]}) \\
			\mathcal{H}_{L}(y^{\mathrm{d},1}_{[0,T_{1}-1]}) & \cdots & \mathcal{H}_{L}(y^{\mathrm{d},q}_{[0,T_{q}-1]})
		\end{bmatrix} g
	\end{equation}
	for some real vector $g\in \mathbb{R}^{\sum_{i=1}^{q} (T_{i}-L+1)}$. 
\end{lemma}

To satisfy the collectively persistently exciting condition in Lemma~\ref{lemma_mosaic}, one requires that $\sum_{i=1}^{q} T_{i}\ge (m+q)(L+n)-q$ and $T_{i}\ge L+n$. The minimum-dimension version of Lemma~\ref{lemma_mosaic} can be concluded, as follows:

\begin{lemma}\label{lemma_mosaic_minimum}
	Consider a controllable LTI system \eqref{equ_LTI} and assume that the input sequences $u^{\mathrm{d},i}_{[0,T_{i}-1]}$ are collectively persistently exciting of order $n+L$. 
	The following statements hold:
	\begin{enumerate}
		\item The rank of the mosaic-Hankel matrix
		\begin{equation} \label{equ_inputoutputMosaic}
			\begin{bmatrix}
				\mathcal{H}_{L}(u^{\mathrm{d},1}_{[0,T_{1}-1]}) & \cdots & \mathcal{H}_{L}(u^{\mathrm{d},q}_{[0,T_{q}-1]}) \\
				\mathcal{H}_{L}(y^{\mathrm{d},1}_{[0,T_{1}-1]}) & \cdots & \mathcal{H}_{L}(y^{\mathrm{d},q}_{[0,T_{q}-1]})
			\end{bmatrix}
		\end{equation}
		satisfies
		\begin{equation} \label{eq:rank-of-mosaic}
			\begin{aligned}
				r &:= \mathrm{rank}\left(\begin{bmatrix}
					\mathcal{H}_{L}(u^{\mathrm{d},1}_{[0,T_{1}-1]}) & \cdots & \mathcal{H}_{L}(u^{\mathrm{d},q}_{[0,T_{q}-1]}) \\
					\mathcal{H}_{L}(y^{\mathrm{d},1}_{[0,T_{1}-1]}) & \cdots & \mathcal{H}_{L}(y^{\mathrm{d},q}_{[0,T_{q}-1]})
				\end{bmatrix}\right) 
				\\
				& \leq mL + n.
			\end{aligned}
		\end{equation}
		\item Suppose $\bar{\mathcal{H}}^{\mathrm{mosaic}}_{L} \in \mathbb{R}^{(m+p)L \times r}$ has the same range~space with the mosaic-Hankel matrix~\eqref{equ_inputoutputMosaic}. Then, any length-$L$ sequence $(u_{[0, L-1]}, y_{[0,L-1]})$ is an input/output trajectory of \eqref{equ_LTI} if and only if we have
		\begin{equation} \label{eq:mosaic-lemma-minimum}
			\begin{bmatrix}
				u_{[0,L-1]} \\ y_{[0,L-1]}
			\end{bmatrix} = \bar{\mathcal{H}}^{\mathrm{mosaic}}_{L} \bar{g}
		\end{equation}
		for some real vector $\bar{g} \in \mathbb{R}^{r}$. 
	\end{enumerate}
\end{lemma}

By the definition of mosaic-Hankel matrix, we have 
\begin{equation*}
	\begin{aligned}
		&\begin{bmatrix}
			\mathcal{H}_{L}(u^{\mathrm{d},1}_{[0,T_{1}-1]}) & \cdots & \mathcal{H}_{L}(u^{\mathrm{d},q}_{[0,T_{q}-1]}) \\ \hline
			\mathcal{H}_{L}(y^{\mathrm{d},1}_{[0,T_{1}-1]}) & \cdots & \mathcal{H}_{L}(y^{\mathrm{d},q}_{[0,T_{q}-1]})
		\end{bmatrix} 
		\\
		&= 
		\begin{bmatrix}
			I_{mL} & 0_{mL\times n} \\ \hline
			\mathcal{T}_{L} & \mathcal{O}_{L}
		\end{bmatrix} 
		\begin{bmatrix}
			\mathcal{H}_{L}(u^{\mathrm{d},1}_{[0,T_{1}-1]}) & \cdots & \mathcal{H}_{L}(u^{\mathrm{d},q}_{[0,T_{q}-1]}) \\ \hline
			\mathcal{H}_{1}(x^{\mathrm{d},1}_{[0,T_{1}-L]}) & \cdots & \mathcal{H}_{1}(x^{\mathrm{d},q}_{[0,T_{q}-L]})
		\end{bmatrix}.
	\end{aligned}
\end{equation*}
The collectively persistently exciting condition ensures that 
\[
\mathrm{rank}\left( \begin{bmatrix}
	\mathcal{H}_{L}(u^{\mathrm{d},1}_{[0,T_{1}-1]}) & \cdots & \mathcal{H}_{L}(u^{\mathrm{d},q}_{[0,T_{q}-1]}) \\ \hline
	\mathcal{H}_{1}(x^{\mathrm{d},1}_{[0,T_{1}-L]}) & \cdots & \mathcal{H}_{1}(x^{\mathrm{d},q}_{[0,T_{q}-L]})
\end{bmatrix} \right) = mL+n.
\]
By following the similar arguments in Lemma~\ref{lemma_minimum}, the statements in Lemma~\ref{lemma_mosaic_minimum} can be obtained.

\section{Proof of Theorem~\ref{theorem1}}
We here present the proof details of Theorem 1. Since $\mathcal{U}$ and $\mathcal{Y}$ are convex polytopes, we can rewrite $u\in \mathcal{U}$ and $y\in \mathcal{Y}$ as $B\begin{bmatrix}
	u^\tr, y^\tr
\end{bmatrix}^\tr \le c$ for some matrix $B$ and vector $c$. Both \eqref{equ_DeePC} and \eqref{equ_DeePC_new} have unique optimal solutions since they are strongly convex. 
Eliminating the variables $u, y, \sigma_{u}, \sigma_{y}$ in the DeePC \eqref{equ_DeePC} leads to
\begin{equation} \label{equ_DeePC_g}
	{
		\begin{aligned} 
			\min_{g} & \quad  \left\|\mathcal{H}_{L}g - b\right\|_{P}^{2} + \lambda_{g}\left\|g \right\|_{2}^{2}
			\\
			\,\, \mathrm{subject~to} &\quad  
			Cg \le c,
		\end{aligned}
	}
\end{equation}  
where $b=\begin{bmatrix}
	u_{\mathrm{ini}}^{\tr}, 0, y_{\mathrm{ini}}^{\tr}, y_{r}^{\tr}
\end{bmatrix}^{\tr}$, $P$ is the block-diagonal matrix $\mathrm{blkdiag}(\lambda_{u}I_{mT_{\mathrm{ini}}}, R, \lambda_{y}I_{pT_{\mathrm{ini}}}, Q)$, and $C=B\begin{bmatrix}
	U_{f}^{\tr}, Y_{f}^{\tr}
\end{bmatrix}^{\tr}$. Based on \eqref{equ_DeePC_g}, we define the Lagrangian $\mathcal{L}(g, \mu) = \left\|\mathcal{H}_{L}g - b\right\|_{P}^{2} + \lambda_{g}\left\|g \right\|_{2}^{2} + \mu^{\tr}(Cg-c)$, where $\mu$ denotes the dual variable. Let $(g^{*}, \mu^{*})$ be the minimizer of \eqref{equ_DeePC_g}. Then, it should satisfy the following KKT condition:
\begin{equation} \label{equ_KKT}
	\left\lbrace
	\begin{aligned}
		&2\mathcal{H}_{L}^{\tr}P(\mathcal{H}_{L}g^{*}-b)+2\lambda_{g}g^{*}+C^{\tr}\mu^{*} = 0,\\
		&\mu^{*\tr}(Cg^{*}-c) = 0, \\
		&Cg^{*}\le c, \quad \mu^{*} \ge 0.
	\end{aligned} 
	\right. 
\end{equation}
Similarly, \eqref{equ_DeePC_new} can be reformulated as
\begin{equation} \label{equ_DeePC_g_new}
	\begin{aligned} 
		\min_{\bar{g}} \quad &  \left\|\bar{\mathcal{H}}_{L}\bar{g} - b\right\|_{P}^{2} + \lambda_{g}\left\| \bar{g} \right\|_{2}^{2}
		\\
		\mathrm{subject~to} \quad  &
		\bar{C}\bar{g} \le c,
	\end{aligned}
\end{equation}
where $\bar{C} = CV_{1}$. The Lagrangian of \eqref{equ_DeePC_g_new} is in the form of $\bar{\mathcal{L}}(\bar{g}, \bar{\mu}) = \left\|\bar{\mathcal{H}}_{L}\bar{g} - b\right\|_{P}^{2} + \lambda_{g}\left\|\bar{g} \right\|_{2}^{2} + \bar{\mu}^{\tr}(\bar{C}\bar{g}-c)$, where $\bar{\mu}$ denotes the dual variable. By utilizing $\bar{\mathcal{H}}_{L} = \mathcal{H}_{L}V_{1}$ (i.e., \eqref{eq:newHL}), $\bar{C} = CV_{1}$ and $V_{1}^{\tr}V_{1}=I_{r}$, the KKT condition of \eqref{equ_DeePC_g_new} is
\begin{equation} \label{equ_KKT_new}
	\left\lbrace
	\begin{aligned}
		&V_{1}^{\tr} (2\mathcal{H}_{L}^{\tr}P(\mathcal{H}_{L}V_{1} \bar{g}^{*}-b)+2\lambda_{g}V_{1} \bar{g}^{*}+C^{\tr} \bar{\mu}^{*}) = 0,\\
		&\bar{\mu}^{*\top}(CV_{1}\bar{g}^{*}-c) = 0, \\
		&CV_{1} \bar{g}^{*}\le c, \quad \bar{\mu}^{*} \ge 0.
	\end{aligned}
	\right. 
\end{equation}
From \eqref{eq:SVD} and $C=B\begin{bmatrix}
	U_{f}^{\tr}, Y_{f}^{\tr}
\end{bmatrix}^{\tr}$, we know that $V_{1}V_{1}^{\tr}=I_{T-L+1}-V_{2}V_{2}^{\tr}$, $V_{1}^{\tr}V_{2}=0$, $\mathcal{H}_{L}V_{2}=0$, and $CV_{2}=0$. Note that \eqref{equ_KKT_new} and these matrix properties are upheld by requiring $\bar{\mathcal{H}}_{L}$, generated from \eqref{eq:newHL}, has the same range space as $\mathcal{H}_{L}$.

Based on the aforementioned matrix properties and \eqref{equ_KKT}, it is easy to verify that $(\bar{g}^{*}=V_{1}^{\tr}g^{*}, \bar{\mu}^{*}=\mu^{*})$ satisfies the KKT condition \eqref{equ_KKT_new}. Therefore, if $g^{*}$ is the minimizer of \eqref{equ_DeePC_g} (i.e., \eqref{equ_DeePC}), then $\bar{g}^{*}=V_{1}^{\tr}g^{*}$ minimizes $\eqref{equ_DeePC_g_new}$ (i.e., \eqref{equ_DeePC_new}).
Finally, given $\bar{g}^{*}=V_{1}^{\tr}g^{*}$, we have 
$$\bar{\mathcal{H}}_{L}\bar{g}^{*} = \mathcal{H}_{L}(V_{1}V_{1}^{\tr})g^{*} = \mathcal{H}_{L}(I_{T-L+1}-V_{2}V_{2}^{\tr})g^{*} = \mathcal{H}_{L}g^{*},$$ which indicates that \eqref{equ_DeePC} and \eqref{equ_DeePC_new} have the same unique optimal solution $(u^*, y^*, \sigma_{u}^*, \sigma_{y}^*)$.